\documentclass[twocolumn,showpacs,preprintnumbers,amsmath,amssymb,prb]{revtex4}

\usepackage{graphicx}% Include figure files
\usepackage{dcolumn}% Align table columns on decimal point
\usepackage{bm}% bold math
\usepackage{amsmath}

\begin{document}

\title{Sub-microsecond correlations in photoluminescence from InAs quantum dots}

\author{Charles Santori}
 \email{chars@stanford.edu}
 \altaffiliation[Also at ]{Institute of Industrial Science, University of
Tokyo, 4-6-1 Komaba, Meguro-ku, Tokyo 153-8904, Japan}
\author{David Fattal}
\author{Jelena Vu\v{c}kovi\'{c}}
 \altaffiliation{Department of Electrical Engineering,
       Stanford University, Stanford, CA 94305}
\author{Glenn S. Solomon}
 \altaffiliation[Also at ]{Solid-State Photonics Laboratory, Stanford University.}
\author{Edo Waks}
\author{Yoshihisa Yamamoto}
 \altaffiliation[Also at ]{NTT Basic Research Laboratories, Atsugishi, Kanagawa, Japan.}
\affiliation{
Quantum Entanglement Project, ICORP, JST, E.L.~Ginzton Laboratory, Stanford University,
    Stanford, California 94305
}

\date{\today}

\begin{abstract}
Photon correlation measurements reveal memory effects in the
optical emission of single InAs quantum dots with timescales
from 10 to 800~ns.  With above-band optical excitation, a
long-timescale negative correlation (antibunching) is observed,
while with quasi-resonant excitation, a positive
correlation (blinking) is observed.  A simple model based on
long-lived charged states is presented that approximately
explains the observed behavior, providing insight into the
excitation process.
Such memory effects can limit the internal efficiency
of light emitters based on single quantum dots, and could
also be problematic for proposed quantum-computation schemes.
\end{abstract}


\maketitle

\section{Introduction}

A variety of memory effects have been reported in the optical emission
of single semiconductor quantum dots.~\cite{generaldot}  Many of these
effects occur on millisecond timescales, including
blinking,~\cite{diff_efros,diff_pistol,diff_sugisaki}
two-color blinking,~\cite{diff_bertram}
and spectral diffusion.~\cite{diff_neuhauser,diff_robinson}
Evidence suggests that millisecond blinking, seen only in
a small minority of quantum dots, is caused by nearby
defects.~\cite{diff_sugisaki}  However, we have previously reported a
much faster type of blinking that occurs in a majority of quantum
dots subject to resonant optical excitation.~\cite{sps_santori}
This blinking behavior appears as a positive correlation in
two-photon coincidence measurements.  Blinking has
also been observed in diamond color centers,~\cite{sps_beveratos}
semiconductor nanocrystals,~\cite{nano_kuno} and
in molecules.~\cite{blink_molecules}  In our case, the
correlation timescale depends on the laser excitation power, but can
vary from less than 10~ns to at least 800~ns.

This article presents a detailed study of these fast memory
effects.  In addition to presenting data on the
blinking of quantum dots under resonant excitation, we report
for the first time negative photon correlations
with timescales greater than 100~ns for quantum dots under pulsed
excitation with the laser tuned above the bandgap of the host
semiconductor.  From a physical viewpoint, studying these
complimentary memory effects can provide valuable information about
the states of a quantum dot, the transitions between them, and especially
the nature of the optical excitation process.  From a practical
viewpoint, blinking effects reduce the efficiency of
quantum-optical devices based on single dots,
\cite{sps_santori,sps_michler2,sps_zwiller,sps_moreau,sps_pelton,twop_santori,sps_yuan,sps_vuckovic}
and also bring into question whether quantum dots are stable
enough for use in proposed quantum computation schemes that
involve optical control.
\cite{qcex_imamoglu,qcex_troiani}  Understanding
the mechanisms responsible for these effects is a first
step toward being able to suppress them.

Both negative and positive correlations over
long timescales can be explained reasonably well through a simplified
model presented later in this article.  Memory effects imply multiple
long-lived configurations of the quantum dot, and our analysis suggests that
these are likely states with differing total charge.
As explained below, negative correlations occur because
above-band excitation injects electrons and holes into the dot separately, whereas
positive correlation (blinking) occurs because resonant excitation
injects electrons and holes together, in pairs.

\section{Samples}

Sample~A, the principle sample used in this study, has been
described in Refs.~\onlinecite{twop_santori,sps_vuckovic}.  It
contains self-assembled InAs quantum dots
(about 25~$\mu$m$^{-2}$) embedded in the middle of a GaAs spacer
layer, and sandwiched between GaAs/AlAs
distributed-Bragg-reflector (DBR) mirrors, grown by molecular-beam
epitaxy.  The quantum dots were grown at a relatively high temperature,
which leads to intermixing between the InAs and surrounding GaAs,
shortening the emission wavelength to approximately
900-950~nm.  Pillars (Fig.~1a) with diameters ranging from
0.3~$\mu$m to 5~$\mu$m and heights of 5~$\mu$m were
fabricated in a random distribution by chemically assisted
ion beam etching (CAIBE) using sapphire dust particles as etch masks.
The resulting microcavities, exhibiting three-dimensional
photon confinement,~\cite{cav_gerard,cav_solomon}
have quality factors of approximately 1000 and spontaneous-emission
rate enhancement (Purcell) factors as high as~5.
The purpose of the optical microcavity was to enhance the
photon collection efficiency and to decrease the spontaneous
emission lifetime through the Purcell effect.  For this
study, the enhanced collection efficiency was valuable, since
the data collection rate in photon correlation measurements
is proportional to the square of the efficiency.

Sample~B, described in Ref.~\onlinecite{sps_santori}, provided
additional data for Figs.~\ref{f_sats} and~\ref{f_resfits}
below.  This sample contained
quantum dots (about 11~$\mu$m$^{-2}$) embedded in simple mesa
structures (0.2-0.4 $\mu$m diameter) without optical cavities.
The emission wavelengths of these dots were shorter (860-900nm).

\section{Experiment}
\begin{figure}
\includegraphics[width=2.2in]{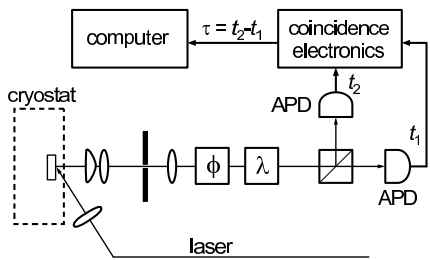}
\caption{\label{f_setup} Schematic diagram of photon correlation
setup.  $\phi$: adjustable half-wave plate followed by a polarizer.
$\lambda$: spectral filter.
}
\end{figure}

The main features of the experimental setups used
for acquiring photon correlation data are shown in Fig.~\ref{f_setup}.
This type of setup, known as a Hanbury Brown and Twiss (HBT)
setup,~\cite{hbt_old} has become a common tool for
studying the dynamics of single quantum systems, including
quantum dots.  This measurement technique can be used to
characterize the performance of single-photon
devices~\cite{sps_santori,sps_michler2,sps_zwiller,sps_moreau,sps_pelton,twop_santori,sps_yuan,sps_vuckovic}
and is more generally useful in studying
spectral emission lines and determining how they are
connected.~\cite{xc_moreau,xc_regelman2,xc_kiraz,polcor_santori,polcor_shields}
For the measurements presented below, several setups were
used, but in all cases the samples were held in a cryostat at
temperatures ranging from 4-15K and excited from a steep
angle (about 54$^\circ$) by 3~ps pulses every
$T_{\rm rep}=13\,{\rm ns}$
from a modelocked Ti-Sapphire laser.  The emission was
collected by a lens (NA=0.5), and imaged onto a pinhole
to define a collection region, approximately 6~$\mu$m wide, on
the sample.  A single linear polarization was selected by
a half-wave plate followed by a fixed polarizer.  The emission
was then spectrally filtered using a diffraction grating in
a monochromator-type configuration, providing spectral
resolutions from about 0.1~nm to 0.35~nm for the various setups.
The spectral filter allows one to collect just
a single emission line of a quantum dot.

The HBT portion
of the setup consists of a beamsplitter with each output
leading to a photon counter.  The photon counters
were EG\&G SPCM avalanche photodiodes, having
about 200~s$^{-1}$ dark counts.  The timing resolution
varied from about 400~ps to 1.3~ns, depending on how
narrowly the light was focused onto the
detector active areas.  Coincidence electronics, consisting of a
time-to-amplitude converter (TAC) followed by a multi-channel
analyzer (MCA) computer card, generated a histogram
of the relative delay $\tau = t_2 - t_1$ between photon
detections at the two counters ($i=1,2$) at times $t_i$.

\begin{figure}
\includegraphics[width=3in]{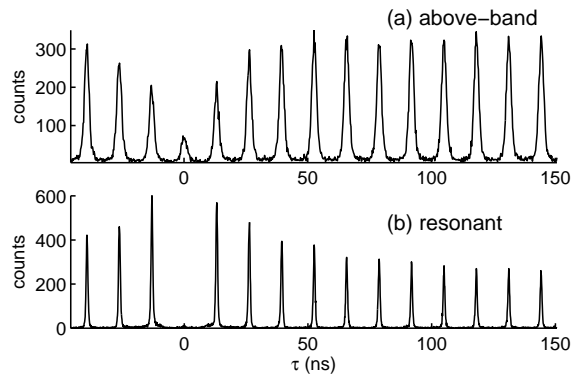}
\caption{\label{f_two_hist} Photon correlation histograms from
quantum dot~1 measured with (a) above-band excitation (750~nm), and
(b) resonant excitation (904~nm).  The difference in the peak widths
between the two cases appears mainly because different setups
with different timing resolutions were used.
}
\end{figure}

\begin{figure}
\includegraphics[width=2.7in]{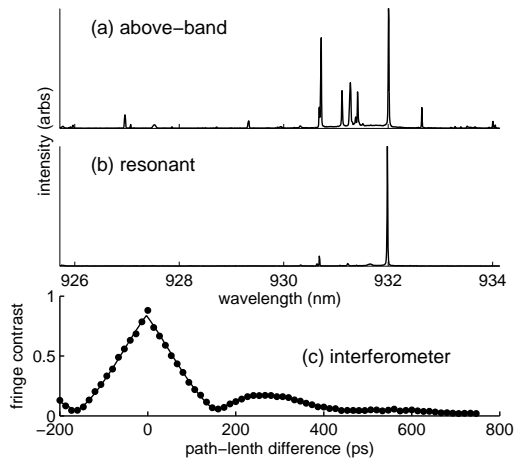}
\caption{\label{f_gspectra} Spectral data for quantum dot~1:
(a) photoluminescence
under above-band excitation (750~nm); (b) photoluminescence under
resonant excitation (909~nm); (c) measurement from a Michelson
interferometer showing fringe contrast vs. path-length difference
for the brightest emission line.
}
\end{figure}

The two types of memory effects we have observed
appear in the photon correlation measurements
shown in Fig.~\ref{f_two_hist}.
The peaks at $\tau = nT_{\rm rep}$ correspond to events
for which one photon was detected from some pulse $m$,
and a second photon was detected from pulse $m+n$.  The
area of the central peak at $\tau=0$ gives information
about photon number statistics within a single pulse.  The
side peaks at $\tau\ne 0$ give information on how
the emission from different pulses is correlated.
Both histograms were obtained from the same quantum
dot on sample~A, dot~1.  The emission was collected
from a bright spectral line at about 932~nm, shown in
Fig.~\ref{f_gspectra}.  In Fig.~\ref{f_two_hist}(a),
the excitation laser was tuned above the bandgap of
the GaAs material surrounding the quantum dot, and the
excitation intensity was chosen so that the collected
emission intensity was far below its maximum value.
The decrease of the side peaks near $\tau = 0$ indicates
a long-term (27 ns) anticorrelation between photons in
consecutive pulses.  This is a new effect not previously
reported.
In Fig.~\ref{f_two_hist}(b), the excitation
laser was resonant with an excited level of the
quantum dot at 904~nm.  Such resonances are found through
a photoluminescence excitation (PLE)
measurement.~\cite{res_kamada,rel_toda,res_finley}
The photoluminescence intensity is monitored as a function
of laser wavelength, and typically 2 or 3 peaks are found.
The rise of the side peaks in the photon correlation
histogram near $\tau = 0$ is opposite from the behavior in
(a), and suggests a blinking of the quantum dot
between a configuration that can emit light at the
wavelength of our spectral filter, and one or more
other configurations that cannot emit at this wavelength
(but could possibly emit at other wavelengths).
The fact that emission from the same quantum dot
can have either positive or negative correlations
over large timescales, depending on the laser
wavelength, implies that there must be an important
difference in how carriers are injected into the
quantum dot for above-band and resonant excitation.

Fig.~\ref{f_gspectra}(a,b) shows photoluminescence spectra from
dot~1.  The measurements in Fig.~\ref{f_two_hist} were performed
on the bright emission line at 932~nm.  In (a), the excitation
was at 750~nm, above the GaAs bandgap, and many emission lines
appear.  Some of these are probably from the same quantum dot
(trion and biexciton lines, for example) while others could be
from other quantum dots.  In (b), resonant excitation at 909~nm
was used, and most of the other peaks have disappeared, demonstrating
the selective nature of resonant excitation.
Fig.~\ref{f_gspectra}(c) shows data obtained by sending light
from the main emission line through a Michelson interferometer.
This setup is described in Ref.~\onlinecite{twop_santori}
and is similar to the one in Ref.~\onlinecite{linw_kammerer2}.
This measurement, performed without
a polarizer, reveals fine structure though an oscillation or
``beating'' in the interference fringe contrast as the path
length is varied, indicating that the emission line is actually
a doublet with a 13~$\mu$eV splitting.  Further measurements
have shown that the two components have orthogonal
linear polarizations, as in Ref.~\onlinecite{ex_kulakovskii}.
The existence of such a splitting suggests that this line
originates from a neutral-exciton transition, rather than
a charged-exciton (trion) transition, which would be
polarization-degenerate in the absence of a magnetic
field.~\cite{ex_bayer}
For the measurements in Fig.~\ref{f_two_hist}, just one
component of the doublet was used, selected through
polarization.

\begin{figure}
\centering
\includegraphics[width=3.4in]{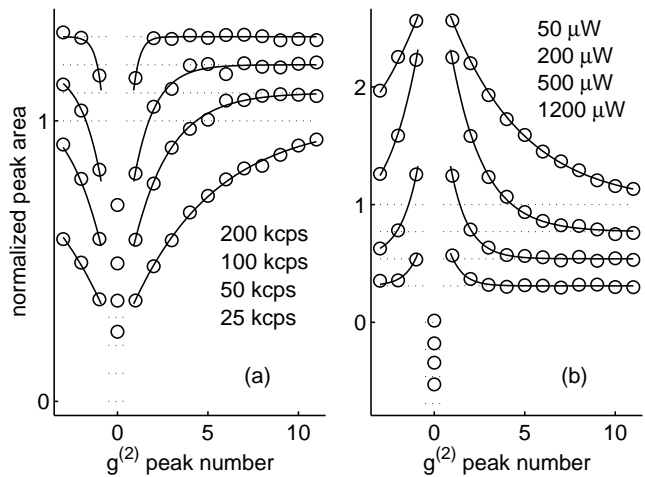}
\caption{\label{f_blinkfits} Normalized peak areas from
photon correlation histograms plotted vs. peak number
(circles), and two-sided exponential fits (lines) using
Eq.~\ref{eq_twoexp}.
(a) Results for dot~1 under above-band excitation, measured
at four different excitation powers, resulting in the
indicated count rates on one detector.  The data sets are shifted
vertically for clarity.
(b) Results for dot~2 under resonant excitation, measured
at four different excitation powers, as indicated.
}
\end{figure}

Several photon correlation measurements taken at different
excitation powers are shown in Fig.~\ref{f_blinkfits}.
In this figure, only the normalized areas of the peaks from
the measured histograms are shown,
plotted versus peak number, with peak 0 at $\tau=0$.
For an ideal $g^{(2)}$ measurement, the normalized peak areas
can be written as,
\begin{equation}
\label{eq_normg2}
g^{(2)}[n] = \frac{\langle n_1[m] n_2[m+n] \rangle}
             {\langle n_1[m] \rangle\langle n_2[m] \rangle} \, ,
\end{equation}
where $n_i[m]$ is the number of photons measured on detector
$i$ from pulse $m$.  The light source is assumed to be stationary:
a shift in $m$ does not change the expectation values in the
numerator or denominator.  The histogram peak areas
obtained from our setup are
approximately proportional to the numerator,
as long as the mean count rates are small compared with the inverses
of the relevant dead times ($\approx 50 \, {\rm ns}$ for the
photon counters, $\approx 1 \, \mu {\rm s}$ for the
electronics).  The effects of these dead times can also be corrected
through calibration relative to scattered laser light.
The denominator can be calculated from the mean
count rates on the detectors.

For the quantum dots we have studied, when absolute normalization
using measured count rates was performed, the peak areas
$g^{(2)}[n]$ converged to $1$ as $n\rightarrow\infty$, and
were well fit by a simple two-sided exponential function:
\begin{equation}
\label{eq_twoexp}
g^{(2)}[n \ne 0] = 1 + g_1 \exp [ - (|n|-1)T_{\rm rep} / \tau_b ] \, ,
\end{equation}
where $g_1$ and $\tau_b$ are fitting parameters that characterize
the amplitude and timescale of the memory effect, respectively.
This shows that there were no additional blinking effects occurring
with timescales greater than $\tau_b$, up to $\approx 100\,{\rm s}$.
The fits using Eq.~\ref{eq_twoexp} typically have
errors that can be explained in terms of statistical Poisson
$\sqrt{N}$ fluctuations in the measured peak areas.
As discussed below, this implies
that a two-state Markov process is sufficient to describe the
observed memory effects.
When absolute normalization was not possible (due
insufficient count-rate data), relative normalization was
performed by fitting Eq.~\ref{eq_twoexp} to the data with
an additional fitting parameter, a normalization coefficient
multiplied by the entire right-hand side.
This procedure was necessary for the data
in Fig.~\ref{f_blinkfits}, while absolute
normalization was possible for the data from dots~3,
4, 6, and~7 in Figs.~\ref{f_abfits} and~\ref{f_resfits},
presented below.

Both the negative and positive correlation
effects in Fig.~\ref{f_blinkfits} display a strong
dependence on laser excitation power.  The four
measurements in Fig.~\ref{f_blinkfits}(a) were taken
from dot~1 under above-band excitation, while the excitation
power was varied to produce mean count rates on one photon counter
of 25, 50, 100, and 200 kcps (kilocounts/second),
as indicated.  The measurements in Fig.~\ref{f_blinkfits}(b)
were taken from dot~2, another quantum dot on sample~A
(results from this dot
appear in Ref.~\onlinecite{sps_vuckovic}).
The emission wavelength was 920~nm, and the excitation
laser was tuned to a resonance at 905~nm.  Four excitation
powers were used, 50, 200, 500, and 1200~$\mu$W, as
indicated.  For both data sets, the timescale of the memory
effect clearly becomes shorter as the excitation power
increases.  This suggests that the fluctuations are
primarily optically induced, at least for the excitation powers
used here.

\begin{figure}
\includegraphics[width=3in]{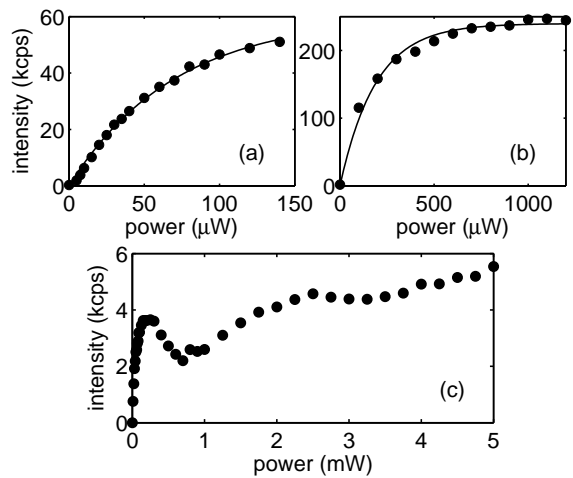}
\caption{\label{f_sats} Mean detector count rate vs.
excitation power (circles), plotted for:
(a) dot~3 under pulsed above-band excitation;
(b) dot~2 under pulsed resonant excitation; (c) a quantum
dot on sample~B under pulsed resonant excitation, exhibiting
coherent excitation behavior.  The
curves in (a) and (b) are fits using
Eq.~\ref{eq_empsat}, appropriate for incoherent excitation.
}
\end{figure}

To enable comparisons between different quantum dots and
with theoretical models, the laser excitation powers must
be normalized in a useful way.  When the excitation intensity is
weak, the injection rate of carriers into the dot is expected to be
proportional to intensity.  However, the efficiency
of this process depends on many experimental variables, such as
the size of the focused laser spot, the size and shape of the
pillar structure containing the dot, and the wavelength of the
laser.  The collected emission
intensity also depends on many factors.  In Fig.~\ref{f_sats},
the emission intensity from three quantum dots under pulsed
excitation is plotted as a function of excitation power.
In Fig.~\ref{f_sats}(a), dot~3, another quantum dot on sample~A
emitting at 919~nm, was excited above-band at 750~nm.
In Fig.~\ref{f_sats}(b), dot~2 was excited on resonance
at 905~nm.  In both cases, a saturation behavior is
observed that is well fit by an empirical model,
\begin{equation}
\label{eq_empsat}
I / I_0 = 1 - e^{-P/P_0} \, ,
\end{equation}
where $I$ is the measured intensity, $P$ is excitation power,
and $I_0$ and $P_0$ are constants characterizing the saturation
intensity and power, respectively.  The normalized
excitation power is $p \equiv P/P_0$.  This
behavior is typical of an ``incoherent'' excitation process.
On the other hand, a simple two-level system without dephasing
excited on resonance is expected to undergo Rabi oscillations,
characteristic of ``coherent'' excitation.  This behavior
can in fact occur in quantum dots when an isolated sharp resonance
with little background can be found in the PLE
spectrum.~\cite{rabi_kamada1,rabi_htoon}  Partially coherent
behavior is shown in Fig.~\ref{f_sats}(c) for a quantum dot
on sample~B emitting
at 877~nm and excited at 864~nm.  Oscillations in the emission
intensity are seen as the laser power is increased.  However,
the quantum dots discussed elsewhere in this article did
not show this coherent effect when excited at the chosen
PLE resonances.

Finally, Figs.~\ref{f_abfits} and~\ref{f_resfits} summarize
the behavior of the memory timescale $\tau_b$ and amplitude
$g_1$, both defined in Eq.~\ref{eq_twoexp}, as a function
of normalized excitation power for
several quantum dots.  In Fig.~\ref{f_abfits}, dot~3 was
excited at 750~nm, above the GaAs bandgap.
Fig.~\ref{f_abfits}(a) is a log-log plot showing the fitted
value of $\tau_b / T_{\rm rep}$ as a function of $p$.
The points approximately follow a line of slope $-1$, indicating
that the blinking rate $1/\tau_b$ is approximately proportional
to the excitation power.  The memory timescale can
be as long as 130~ns for small excitation powers, and less than
one pulse period for large powers.  For large enough
powers, the memory timescale becomes so short
that even the innermost peak areas in the photon correlation
histogram have areas close to $1$, and it becomes difficult
to extract a value for $\tau_b$.
Fig.~\ref{f_abfits}(b) is a semilog plot showing
the amplitude $g_1$ as a function of
excitation power.  Fig.~\ref{f_resfits} shows similar data
for dots~2, 4, 5, 6, and 7 under resonant excitation.
Various parameters of these quantum dots are summarized in
Table~\ref{dot_tab}.
Dots~2, 4, 5, and 6 are from sample~A, while dot~7 is from
sample~B.
The blinking rate $1/\tau_b$ is again approximately
proportional to excitation power below the saturation
regime ($P/P_0 < 1$).  The memory timescale was as long
as 770~ns for dot~4 under weak excitation power.
The blinking amplitude $g_1$ shows similar behavior as
with above-band excitation, except that
it is positive.  The curve fits in these two
figures are based on the theoretical model to be introduced
next.
\begin{figure}
\includegraphics[width=3in]{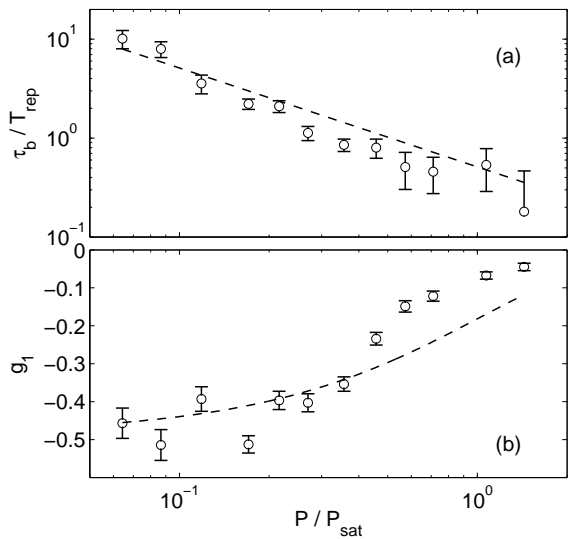}
\caption{\label{f_abfits} Summary of results for dot~3 under
above-band excitation.  Error bars include
only Poisson $\sqrt{N}$ fluctuations in the correlation
peak areas.
(a) Log-log plot showing the fitted blinking
timescale $\tau_b$ divided by $T_{\rm rep}=13\,{\rm ns}$, plotted
vs. normalized laser power $P/P_0$.  The line fit used
Eq.~\ref{eq_finalmodl_a} with $C_1 = -0.96$.
(b) Semilog plot showing blinking amplitude $g_1$ vs. normalized
laser power.  The drawn curve
used Eq.~\ref{eq_finalmodl_b} with $C_2 = 0.51$.
}
\end{figure}
\begin{figure}
\includegraphics[width=3in]{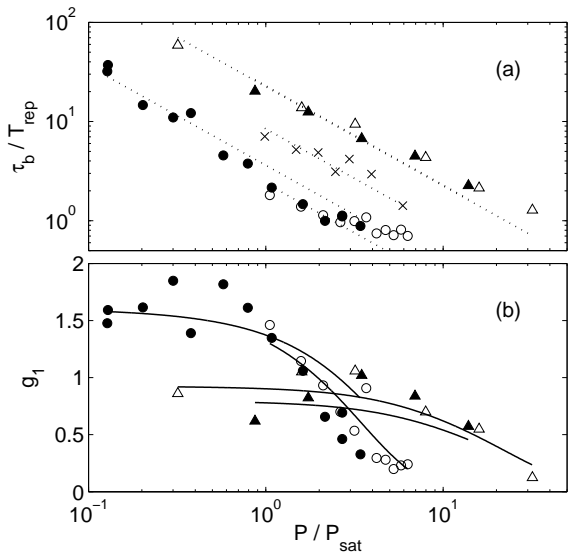}
\caption{\label{f_resfits}
Summary of results for dots~2 (empty circles),
4 (empty triangles), 5 (filled triangles),
6 (filled circles), and 7 (x's) under resonant excitation.
(a) Log-log plot showing the fitted blinking
timescale $\tau_b$ divided by $T_{\rm rep}=13\,{\rm ns}$, plotted
vs. normalized laser power $P/P_0$.  The line fit used
Eq.~\ref{eq_finalmodl_a} with the $C_1$ values
in Table~\ref{dot_tab}.
(b) Semilog plot showing blinking amplitude $g_1$ vs. normalized
laser power.  The drawn curves used
Eq.~\ref{eq_finalmodl_b} with the $C_2$ values
in Table~\ref{dot_tab}.
}
\end{figure}

\section{Theoretical Model}
The simple exponential decays in the measured photon
correlation histograms (Fig.~\ref{f_blinkfits}, for
example) suggest that the memory effects can be
described by a two-state Markov process. In such a
process, only the most recent state of a system is
needed to determine its future evolution.  Suppose
a quantum dot has two stable states $1$ and $2$
in which it can remain long after an excitation pulse.
The effect of a single excitation pulse is described by:
\begin{equation}
\label{eq_markov}
\left( \begin{array}{c} p_1 \\ p_2 \end{array} \right)
\rightarrow
\left( \begin{array}{cc} 1-a & b \\ a & 1-b \end{array} \right)
\left( \begin{array}{c} p_1 \\ p_2 \end{array} \right)
\, ,
\end{equation}
where $p_i$ is the probability to be in state $i$,
and $a$ and $b$ are the $1\rightarrow 2$ and
$2\rightarrow 1$ transition probabilities, respectively.
We next assume that, if we have detected a photon immediately
after an excitation pulse, the system must have ended in
state $1$.  This is reasonable in our experiment, since
we spectrally select an emission line corresponding to
a unique transition.  If a photon was detected from pulse
$0$, the system evolves according to,
\begin{equation}
\label{eq_markev}
\left( \begin{array}{c} p_1[m] \\ p_2[m] \end{array} \right)
=
\frac{1}{a+b}\left( \begin{array}{c} b \\ a \end{array} \right)
+
\frac{a(1-a-b)^m}{a+b}
\left( \begin{array}{c} 1 \\ -1 \end{array} \right) \, ,
\end{equation}
where $p_i[m]$ is the probability of the system to be in state
$i$ after pulse $m$.  Finally, let $\eta_i$ be the
probability of emitting a photon immediately after
an excitation pulse, given the dot was in state $i$ before the
pulse.  One can then calculate $g^{(2)}[n]$, obtaining
a two-sided exponential function as in Eq.~\ref{eq_twoexp}
with parameters,
\begin{subequations}
\label{eq_markovparms}
\begin{eqnarray}
\tau_b  &=& \frac{-T_{\rm rep}}{\ln(1-a-b)} \, ,\\
g_1     &=& \frac{a(\eta_1 - \eta_2)}{b \eta_1 + a \eta_2} \, .
\end{eqnarray}
\end{subequations}
If $\eta_1 > \eta_2$, $g_1$ is positive, as
observed experimentally
with resonant excitation.  If $\eta_1 < \eta_2$, $g_1$
is negative, as observed with above-band excitation.
Another important parameter is the internal efficiency,
that is, the probability of emitting
a photon after any given excitation pulse.  The result is,
\begin{equation}
\label{eq_efftot}
\eta_{\rm tot} = \frac{b \eta_1 + a \eta_2}{a + b} \, .
\end{equation}

We next consider the nature of states $1$ and
$2$.  They can have lifetimes approaching 1$\mu$s,
so they cannot be optically active
states such as excitons, biexcitons, and trions, which
have nanosecond lifetimes.
For quantum dot~1 discussed above, spectroscopy suggests
that the studied emission line is a neutral-exciton transition,
and thus state $1$ is the neutral ground state (empty dot).
For state $2$, the ``dark'' state,
recent spectroscopy literature on self-assembled quantum
dots~\cite{dark_bayer,spec_hartmann,spec_finley,ex_bayer}
and chemically synthesized nanocrystals~\cite{nanocrystal1} suggests two
main possibilities: a ``dark exciton'' or a charged state.
Dark excitons are electron-hole pairs with spins oriented
such that optical recombination is forbidden.  Although
these states can have microsecond lifetimes in
chemically synthesized nanocrystals with typical
diameters less than 5~nm, it is unlikely that they could
be as long-lived in our self-assembled quantum dots.
Self-assembled dots can have reduced symmetry,
so that the dark-exciton transition is not entirely
forbidden.~\cite{ex_bayer}  Additionally, if the energy
splitting between the bright and dark exciton levels is
proportional to the inverse cube of the dot radius,~\cite{nanocrystal1}
spin flip processes might occur much faster in our samples,
with typical dot radii of 30~nm.
Dark excitons in InAs quantum dots have not yet been adequately
studied, and we cannot reject this possibility with certainty.
Nevertheless, we favor the other possibility, that state $2$
is a charged state.  This choice allows one to
explain all of the observed memory effects with a single
model.  Charge fluctuations in quantum dots under
continuous-wave (cw) above-band excitation have been
observed through correlations between exciton and trion
photon emissions,~\cite{xc_kiraz} and we have seen
similar behavior in our samples.  The assumption below
is that the extra charges reside inside the quantum
dot.  An alternative possibility not considered here is
that extra charges could be associated with impurities
outside the quantum dot; this would not explain the
negative correlations observed with above-band excitation.

\begin{figure}
\centering
\includegraphics[width=2.5in]{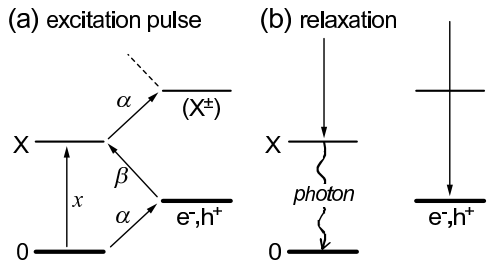}
\caption{\label{f_ladder} Energy levels and transitions
for the blinking model described in the text: (a) a
short period of laser excitation, followed by (b) a
long relaxation period.
}
\end{figure}

The model is shown schematically in Fig.~\ref{f_ladder}.
Level $0$ is the neutral ground state, and level $X$ is the
single-exciton state.  Level $e^-, h^+$ is a charged state.
There could in reality be more than one charged state
involved significantly with the dynamics, but we consider
only one charged state in this model.  Level $X^\pm$ is
a trion (charged-exciton) state.  Some of these levels
can have degeneracies, but in these cases we are interested
only in the total occupation probabilities.

The system evolves in two steps.  First, a strong optical
field is applied for a short time duration $\Delta t$, during
which upward transitions are induced.  Two of these
transitions, with rates $\alpha$ and $\beta$, change the
total charge of the dot from
${\rm neutral} \rightarrow {\rm charged}$ and from
${\rm charged} \rightarrow {\rm neutral}$, respectively.
These transitions correspond physically to the capture of
single electrons and holes from the surrounding region into
the quantum dot.  The third transition, with rate $x$,
brings the quantum dot from $0$ to $X$.  This transition
could occur either through resonant excitation, or through
the capture of an entire electron-hole pair.
During the second step, which
could last several nanoseconds, the system relaxes back
down to levels $0$ and $e^-, h^+$ through electron-hole recombination.
This is assumed to be a charge-conserving process.
If a relaxation
occurs from $X$ to $0$, a photon is emitted at the special
wavelength that our setup detects.

This model can be solved to give the following parameters
for the general Markov process in Eq.~\ref{eq_markov}:
\begin{subequations}
\label{eq_chargemodl}
\begin{eqnarray}
a &=& \frac{\alpha}{\alpha+\beta} (1 - e^{-(\alpha+\beta)\Delta t}) \, , \\
b &=& \frac{\beta}{\alpha+\beta} (1 - e^{-(\alpha+\beta)\Delta t}) \, , \\
\eta_1 &=& 1 - a - e^{-(\alpha+x)\Delta t} \, , \\
\eta_2 &=& b \, .
\end{eqnarray}
\end{subequations}
To fit this model to the experimental data, we make a further
assumption, that the rates $\alpha$, $\beta$, and
$x$ are proportional to the excitation power $p$.  Inserting
Eqs.~\ref{eq_chargemodl} into Eq.~\ref{eq_efftot} gives the
saturation behavior,
\begin{equation}
\label{eq_modlsat}
\eta_{\rm tot} = \frac{\beta}{\alpha+\beta} (1 - e^{-(\alpha+x)\Delta t}) \, .
\end{equation}
Comparing this with the empirical Eq.~\ref{eq_empsat} allows
the assignment $(\alpha+x)\Delta t = P/P_0 = p$.  We then
introduce two fitting constants, $C_1 = (x-\beta)/(x+\alpha)$
and $C_2 = \alpha / \beta$.  Substituting these and
Eqs.~\ref{eq_chargemodl} into Eqs.~\ref{eq_markovparms} gives
the final result:
\begin{subequations}
\label{eq_finalmodl_both}
\begin{eqnarray}
\label{eq_finalmodl_a}
\tau_b  &=& \frac{T_{\rm rep}}{(1-C_1)p} \, ,\\
\label{eq_finalmodl_b}
g_1     &=& C_2 \left( \frac{e^{C_1 p}-1}{e^{p}-1}    \right) \, .
\end{eqnarray}
\end{subequations}

The parameter $C_1$ has special significance, since it
determines whether electrons and holes are more often added
individually or in pairs.  If they are added individually
($C_1 < 0$), Eq.~\ref{eq_finalmodl_b} predicts negative correlation
($g_1 < 0$) in the detected photons.  If they are added more
often in pairs ($C_1 > 0$), Eq.~\ref{eq_finalmodl_b} predicts
positive correlation ($g_1 > 0$).
These behaviors have simple qualitative explanations.
It follows from Eqs.~\ref{eq_normg2} and~\ref{eq_twoexp} that
the quantity $1+g_1$ is proportional to the conditional probability,
given that a photon was detected from pulse $0$, that a second photon will
be detected from pulse $1$.  When a first photon is detected,
the dot is empty immediately afterward.
For single-carrier injection, two injections must occur
before another photon can be emitted, and thus it is
unlikely that another photon will be emitted from pulse $1$
if the injection rate is small.
The opposite argument applies for the injection of entire
electron-hole pairs.  If a photon was emitted from
pulse $0$, it is especially likely that another
photon will be emitted from pulse $1$, since only a single
additional pair needs to be injected.  Otherwise, the
dot could be charged, in which case the quantum dot will
appear dark at the selected wavelength.

The lines in Figs.~\ref{f_abfits}(a) and~\ref{f_resfits}(a) were
obtained by fitting Eq.~\ref{eq_finalmodl_a} to the measured
blinking timescales.
The values of $C_1$ obtained from these fits are summarized in
Table~\ref{dot_tab}.  As expected from the preceding discussion,
$C_1$ is negative for above-band excitation, and positive in
all five cases for resonant excitation.
Substantial variation in the value of
$C_1$ is seen among the different dots under resonant excitation,
however.  $C_1$ is much closer to its maximum value of $1$
for dots~4 and~5, than for dots~2 and~6.  In our model,
$C_1$ characterizes how ``clean'' the resonant excitation
process is, or in other words how rarely extra charges are
added and removed from the quantum dot.  Perhaps the differences
among these dots are related to the emission wavelength.
It is also possible that for one or more of these dots,
the studied emission line could be a trion (charged exciton)
transition.  In those cases our model should still apply
(after changing the labels in Fig.~\ref{f_ladder}), but
it would not be surprising if the values of the fitting
parameters were different.

The curves in Figs.~\ref{f_abfits}(b) and~\ref{f_resfits}(b) were
obtained by fitting Eq.~\ref{eq_finalmodl_b} to the data
using the values of $C_1$ already obtained and using $C_2$ as
a fitting parameter.  The fits do not match the data
perfectly, a sign that the model is too simple.
The saturation regime ($P/P_0 > 1$) is difficult to model
accurately, since a large number of states are involved.
The model does correctly predict that $g_1$ tends to
zero for large excitation powers, as observed in
the data.  The parameter $C_2$ has practical
importance, since according to Eq.~\ref{eq_modlsat}, the
maximum internal efficiency at the selected wavelength is
$\eta_{\rm max} = \beta / (\alpha + \beta) = 1 / (1+C_2)$.
It should be cautioned, however, that using this
formula with the fitted values of $C_2$ may not give
the correct efficiency in the saturation regime, where
the model is least accurate.

An important question is why the charge of a quantum dot should
ever change when the excitation wavelength is tuned below
the GaAs bandgap and below the InAs wetting layer band edge.
This could be related to the ``wetting layer tail''
observed in PLE spectra.  This feature has been attributed
to continuum states associated with the combined wetting
layer-quantum dot system.~\cite{rel_toda}  If these
states are not localized to a single quantum dot, electrons
and holes excited into these states by a laser pulse
could be captured by different quantum dots, changing
their charges, for example.
It has also been suggested that an Auger-type process, which would
allow an electron to escape from the quantum dot, could play
an important role in the relaxation of electron-hole pairs from
excited states.~\cite{auger_norris,auger_sanguinetti}
Two-photon processes are an
unlikely mechanism when the excitation power is far below
saturation.  This is because the blinking rate $1/\tau_b$ is
observed to vary approximately linearly with the excitation
power.
\begin{table}
\caption{\label{dot_tab} Parameters of the studied quantum dots.
$\lambda_0$ and $\lambda_{exc}$ are emission and excitation wavelengths,
respectively.}
\begin{ruledtabular}
\begin{tabular}{lccccccc}
  dot \#              & 1       & 2     & 3     & 4     & 5     & 6    & 
                                         7\footnote[0]{$^*$on Sample B}{*} \\
  $\lambda_0$ (nm)    & 932     & 920   & 919   & 938   & 940   & 921  & 876   \\ 
  $\lambda_{exc}$ (nm)& 750/904 & 905   & 750   & 913   & 904   & 892  & 858   \\
  $P_0$ ($\mu$W)      &  -      & 190   & 70    & 3.1   & 115   & 28   & 446   \\
  $C_1$               &  -      & 0.58  & -0.96 & 0.956 & 0.956 & 0.72 & 0.88  \\
  $C_2$               &  -      & 2.9  & 0.51  & 0.97  & 0.84  & 2.2   &  -    \\
\end{tabular}
\end{ruledtabular}
\end{table}

\section{Conclusions}
We have observed sub-microsecond memory effects in quantum-dot
photoluminescence on two samples, one with optical microcavities and
one without.  Other groups have not yet reported similar
behavior, but other studies have not, to our knowledge,
explored pulsed resonant excitation, or pulsed above-band excitation
far below the saturation level, the two regimes important in
our study.  These memory effects imply the existence
of multiple long-lived states, which are most likely
states with different total charge.  If this interpretation
is correct, it is apparently difficult to prevent the
charge of a semiconductor quantum dot from changing during
optical excitation.

Fluctuations in the charge of a quantum dot could pose
serious difficulties for proposed applications.
Light emitters based on single quantum dots will have reduced
internal efficiencies at a particular wavelength.
Even though the blinking seems to disappear at large excitation
powers in photon correlation measurements,
our model suggests that the efficiency still suffers,
due to the presence of multiple configurations.
Optically induced charge fluctuations could also be
problematic in schemes for quantum computation that
involve optical control of single charges or excitons.

Better stability might be achieved
using laser excitation resonant with the fundamental exciton
transition.  We did not attempt this because of the
difficulty of removing scattered laser light, but at least
one group has performed single-dot measurements
under these conditions.~\cite{spec_bonadeo1}
Another option might be to design the
optical excitation process so that
the transition rate into an unwanted configuration is much
smaller than the transition rate out of it.  In this case,
the quantum dot would spend most of its time in the desired
configuration.  Finally, if charge
fluctuations are, in fact, related to continuum states, then
fabrication of quantum dots with a low density and without
a wetting layer might be advantageous.

\begin{acknowledgments}
Supported in part by MURI UCLA/0160-G-BC575.  The authors
thank A.~Scherer and T.~Yoshie from Caltech for providing
access to CAIBE and for help with fabrication, and M.~Pelton
for patterning of sample~B and for useful comments.
\end{acknowledgments}


\begin{thebibliography}{99}
\bibitem{generaldot}
D. Bimberg, M.~Grundmann, and N.~N. Ledentsov,
{\it Quantum Dot Heterostructures} (John Wiley \& Sons, Chichester, 1999).
%
% spectral diffusion
%
\bibitem{diff_efros}
Al.~L.~Efros and M.~Rosen,
%``Random telegraph signal in the photoluminescence
%intensity of a single quantum dot,''
Phys. Rev. Lett. {\bf 78}, 1110 (1997).
%
\bibitem{diff_pistol}
M-E.~Pistol, P.~Castrillo, D.~Hessman, J.~A.~Prieto, and L.~Samuelson,
%``Random telegraph noise in photoluminescence from individual self-assembled
%quantum dots,''
Phys. Rev. B {\bf 59}, 10725 (1999).
%
\bibitem{diff_sugisaki}
M.~Sugisaki, H.~W.~Ren, K.~Nishi, and Y.~Masumoto,
%``Fluorescence intermittency in self-assembled InP quantum dots,''
Phys. Rev. Lett. {\bf 86}, 4883 (2001). 
%
\bibitem{diff_bertram}
D.~Bertram, M.~C.~Hanna, and A.~J.~Nozik,
%``Two color blinking of single strain-induced GaAs quantum dots,''
Appl. Phys. Lett. {\bf 74}, 2666 (1999).
%
\bibitem{diff_neuhauser}
R.~G.~Neuhauser, K.~T.~Shimizu, W.~K.~Woo, S.~A.~Empedocles, and M.~G.~Bawendi,
%``Correlation between Fluorescence Intermittency and Spectral Diffusion in
%Single Semiconductor Quantum Dots,''
Phys. Rev. Lett. {\bf 85}, 3301 (2000).
%
\bibitem{diff_robinson}
H.~D.~Robinson and B.~B.~Goldberg,
%``Light-induced spectral diffusion in single
%self-assembled quantum dots,''
Phys. Rev. B {\bf 61}, R5086 (2000).
%
%
%
%
\bibitem{sps_santori}
C.~Santori, M.~Pelton, G.~Solomon, Y.~Dale, and Y.~Yamamoto,
%``Triggered single photons from a quantum dot,''
Phys. Rev. Lett. {\bf 86}, 1502 (2001).
%
\bibitem{sps_beveratos}
A.~Beveratos, S.~K\"{u}hn, R.~Brouri, T.~Gacoin,
J.-P.~Poizat, and P.~Grangier,
%``Room temperature stable single-photon source,''
Eur. Phys. J. D, {\bf 18}, 191 (2002).
%
\bibitem{nano_kuno}
M.~Kuno, D.~P.~Fromm, H.~F.~Hamann, A.~Gallagher, and
D.~J.~Nesbitt,
%nonexponential ``blinking'' kinetics of single CdSe quantum
%dots: a universal power law behavior,
J. Chem. Phys. {\bf 112}, 3117 (2000).
%
\bibitem{blink_molecules}
S.~C.~Kitson, P.~Jonsson, J.~G.~Rarity, and P.~R.~Tapster,
Phys. Rev. A {\bf 58}, 620 (1998).
%
%
%quantum dots
%
%
\bibitem{sps_michler2}
P.~Michler, A.~Kiraz, C.~Becher, W.~V.~Schoenfeld, P.~M.~Petroff,
L.~Zhang, E.~Hu, and A.~Imamoglu,
%``A quantum dot single-photon turnstile device,''
Science {\bf 290}, 2282 (2000).
%
\bibitem{sps_zwiller}
V.~Zwiller, H.~Blom, P.~Jonsson, N.~Panev, S.~Jeppesen,
T.~Tsegaye, E.~Goobar, M.-E.~Pistol, L.~Samuelson, and G.~Bj\"{o}rk,
%``Single quantum dots emit single photons at a time:
%antibunching experiments,''
Appl. Phys. Lett. {\bf 78}, 2476 (2001).
%
\bibitem{sps_moreau}
E.~Moreau, I.~Robert, J.-M.~G\'{e}rard, I.~Abram,
L.~Manin, and V.~Thierry-Mieg,
%``Single-mode solid-state single photon source based on isolated
%quantum dots in pillar microcavities,''
Appl. Phys. Lett. {\bf 79}, 2865 (2001).
%
\bibitem{sps_pelton}
M.~Pelton, C.~Santori, J.~Vu\v{c}kovi\'{c}, B.~Zhang,
G.~S.~Solomon, J.~Plant, and Y.~Yamamoto,
%``An efficient source of single photons: a single quantum dot
%in a micropost microcavity,''
Phys. Rev. Lett. {\bf 89}, 233602 (2002).
%
\bibitem{twop_santori}
C.~Santori, D.~Fattal, J.~Vu\v{c}kovi\'{c}, G.~S.~Solomon, and
Y.~Yamamoto,
%``Indistinguishable Photons from a Single-Photon Device,''
Nature (London) {\bf 419}, 594 (2002).
%
\bibitem{sps_yuan}
Z.~Yuan, B.~E.~Kardynal, R.~M.~Stevenson, A.~J.~Shields, C.~J.~Lobo,
K.~Cooper, N.~S.~Beattie, D.~A.~Ritchie, and M.~Pepper,
%``Electrically driven single photon source,''
Science {\bf 295}, 102 (2002).
%
\bibitem{sps_vuckovic}
J.~Vu\v{c}kovi\'{c}, D.~Fattal, C.~Santori, G.~S.~Solomon, and
Y.~Yamamoto,
Appl. Phys. Lett. {\bf 82}, 3596 (2003).
%
%
% Quantum computation using excitons and charges
%
\bibitem{qcex_imamoglu}
A.~Imamoglu, D.~D.~Awschalom, G.~Burkard, D.~P.~DiVincenzo, D.~Loss,
M.~Sherwin, and A.~Small,
%quantum information processing using quantum dot spins and cavity qed
Phys. Rev. Lett. {\bf 83}, 4204 (1999).
%
\bibitem{qcex_troiani}
Filippo Troiani, Ulrich Hohenester, and Elisa Molinari,
%``Exploiting exciton-exciton interactions in semiconductor quantum dots for
%quantum-information processing,''
Phys. Rev. B {\bf 62}, R2263 (2000).
%
%
%
%
% microcavities
%
%
%
\bibitem{cav_gerard}
J.~M.~G\'{e}rard, B.~Sermage, B.~Gayral, B.~Legrand, E.~Costard, and
V.~Thierry-Mieg,
%``Enhanced spontaneous emission by quantum boxes in a
%monolithic optical microcavity,''
Phys. Rev. Lett. {\bf 81}, 1110 (1998).
%
\bibitem{cav_solomon}
G.~S.~Solomon, M.~Pelton, and Y.~Yamamoto,
%``Single-mode spontaneous emission from a single quantum dot in a
%three-dimensional microcavity,''
Phys. Rev. Lett. {\bf 86}, 3903 (2001).
%
%
%HBT references
\bibitem{hbt_old}
F.~T.~Arecchi, M.~Corti, V.~Degiorgio, and S.~Donati,
Opt. Commun. {\bf 3}, 284 (1971).
%
%
%
% other dot HBT measurements
%
%
%
%
\bibitem{xc_moreau}
E.~Moreau, I.~Robert, L.~Manin, V.~Thierry-Mieg, J.~M.~G\'{e}rard, and I.~Abram,
%``Quantum cascade of photons in semiconductor quantum dots,''
Phys. Rev. Lett. {\bf 87}, 183601-1 (2001).
%
\bibitem{xc_regelman2}
D.~V.~Regelman, U.~Mizrahi, D.~Gershoni, E.~Ehrenfreund, W.~V.~Schoenfeld, and
P.~M.~Petroff,
%``Semiconductor quantum dot: A quantum light source of
%multicolor photons with tunable statistics,''
Phys. Rev. Lett. {\bf 87}, 257401-1 (2001).
%
\bibitem{xc_kiraz}
A.~Kiraz, S.~F\"{a}lth, C.~Becher, B.~Gayral, W.~V.~Schoenfeld, P.~M.~Petroff, L.~Zhang, E.~Hu, and A.~\.{I}mamo\u{g}lu,
%``Photon correlation spectroscopy of a single quantum dot,''
Phys. Rev. B {\bf 65}, 161303-1 (2002).
%
\bibitem{polcor_santori}
C.~Santori, D.~Fattal, M.~Pelton, G.~S.~Solomon, and Y.~Yamamoto,
%``Polarization-correlated photon pairs from a single quantum dot,''
Phys. Rev. B {\bf 66}, 045308 (2002).
%
\bibitem{polcor_shields}
R.~M.~Stevenson, R.~M.~Thompson, A.~J.~Shields, I.~Farrer,
B.~E.~Kardynal, D.~A.~Ritchie, and M.~Pepper,
%``Quantum dots as a photon source for passive quantum key encoding,''
Phys. Rev. B {\bf 66}, 081302 (2002).
%
%
%
%
% resonant excitation
%
%
\bibitem{res_kamada}
H.~Kamada, H.~Ando, J.~Temmyo, and T.~Tamamura,
%Excited-state optical transitions of excitons and biexcitons
%in a single InGaAs quantum disk
Phys. Rev. B {\bf 58}, 16243 (1998).
%
\bibitem{rel_toda}
Y.~Toda, O.~Moriwaki, M.~Nishioka, and Y.~Arakawa,
%``Efficient carrier relaxation mechanism in InGaAs/GaAs self-assembled
%quantum dots based on the existence of continuum states,''
Phys. Rev. Lett. {\bf 82}, 4114 (1999).
%
\bibitem{res_finley}
J.~J.~Finley, A.~Lema\^{i}tre, A.~D.~Ashmore, D.~J.~Mowbray,
M.~S.~Skolnick, M.~Hopkinson, and T.~F.~Krauss,
%Excitation and relaxation mechanisms in single In(Ga)As quantum dots
Phys. Stat. Sol. (b) {\bf 224}, 373 (2001).
%
%
%
% interferometer setup
%
%
\bibitem{linw_kammerer2}
C.~Kammerer, G.~Cassabois, C.~Voisin, M.~Perrin, C.~Delalande,
Ph.~Roussignol, and J.~M.~G\'{e}rard,
%``Interferometric correlation spectroscopy in single quantum dots,''
Appl. Phys. Lett. {\bf 81}, 2737 (2002).
%
\bibitem{ex_kulakovskii}
V.~D.~Kulakovskii, G.~Bacher, R.~Weigand, T.~K\"{u}mmell,
A.~Forchel, E.~Borovitskaya, K.~Leonardi, and D.~Hommel,
%``Fine structure of biexciton emission in symmetric and asymmetric
%CdSe/ZnSe single quantum dots,''
Phys. Rev. Lett. {\bf 82}, 1780 (1999).
%
\bibitem{ex_bayer}
M.~Bayer, G.~Ortner, O.~Stern, A.~Kuther, A.~A.~Gorbunov, A.~Forchel,
P.~Hawrylak, S.~Fafard, K.~Hinzer, T.~L.~Reinecke, S.~N.~Walck,
J.~P.~Reithmaier, F.~Klopf, and F.~Sch\"{a}fer,
%``Fine structure of neutral and
%charged excitons in self-assembled In(Ga)As/(Al)GaAs quantum dots,''
Phys. Rev. B {\bf 65}, 195315-1 (2002).
%
%
%
% Rabi
%
\bibitem{rabi_kamada1}
H.~Kamada, H.~Gotoh, J.~Temmyo, T.~Takagahara, and H.~Ando,
%``Exciton Rabi oscillation in a single quantum dot,''
Phys. Rev. Lett. {\bf 87}, 246401-1 (2001).
%
\bibitem{rabi_htoon}
H.~Htoon, T.~Takagahara, D.~Kulik, O.~Baklenov, A.~L.~Holmes,
and C.~K.~Shih,
%interplay of rabi oscillations and quantum interference in
%semiconductor quantum dots
Phys. Rev. Lett. {\bf 88}, 087401 (2002).
%
%
%
% spectroscopy
%
%
\bibitem{dark_bayer}
M.~Bayer, O.~Stern, A.~Kuther, and A.~Forchel,
%Spectroscopic study of dark excitons in InGaAs self-assembled
%quantum dots by a magnetic-field-induced symmetry breaking
Phys. Rev. B {\bf 61}, 7273 (2000).
%
\bibitem{spec_hartmann}
A.~Hartmann, Y.~Ducommun, E.~Kapon, U.~Hohenester, and E.~Molinari,
%``Few-particle effects in semiconductor quantum dots: observation of
%multicharged excitons,''
Phys. Rev. Lett. {\bf 84}, 5648 (2000).
%
\bibitem{spec_finley}
J.~J.~Finley, P.~W.~Fry, A.~D.~Ashmore, A.~Lema\^{i}tre, A.~I.~Tartakovskii,
R.~Oulton, D.~J.~Mowbray, M.~S.~Skolnick, M.~Hopkinson, P.~D.~Buckle,
P.~A.~Maksym,
%``Observation of multicharged excitons and biexcitons in a single
%InGaAs quantum dot,''
Phys. Rev. B {\bf 63}, 161305-1 (2001).
%
%
%
% nanocrystals
%
\bibitem{nanocrystal1}
Al.~L.~Efros, M.~Rosen, M.~Kuno, M.~Nirmal, D.~J.~Norris, and M.~Bawendi,
%Band-edge exciton in quantum dots of semiconductors with a degenerate
%valence band: dark & bright exciton states
Phys. Rev. B {\bf 54}, 4843 (1996).
%
%
%
%
\bibitem{auger_norris}
J.~Urayama, T.~B.~Norris, J.~Singh, and P.~Bhattacharya,
%Observation of photon bottleneck in quantum dot electronic
%relaxation
Phys. Rev. Lett. {\bf 86}, 4930 (2001).
%
\bibitem{auger_sanguinetti}
S.~Sanguinetti, K.~Watanabe, T.~Tateno, W.~Wakaki, N.~Koguchi,
T.~Kuroda, F.~Minami, and M.~Gurioli,
%role of the wetting later in the carrier relaxation in quantum dots
Appl. Phys. Lett. {\bf 81}, 613 (2002).
%
% direct excitation of 1X
%
\bibitem{spec_bonadeo1}
N.~H.~Bonadeo, Gang Chen, D.~Gammon, and D.~G.~Steel,
%``Single quantum dot nonlinear optical spectroscopy,''
Phys. Stat. Sol. (b) {\bf 221}, 5 (2000).
%
\end{thebibliography}
\end{document}